\documentclass[conference]{IEEEtran}
\IEEEoverridecommandlockouts

\usepackage{graphicx} 
\graphicspath{ {./Images/} }
\usepackage{amssymb}
\usepackage{latexsym}
\usepackage{soul}  
\usepackage{xcolor}  

\usepackage{mathtools, cuted}
\usepackage{lipsum, color}
\usepackage[font=small]{caption}

\usepackage{amsmath}

\usepackage{listings}
\usepackage{algorithmic}
\usepackage{textgreek}
\usepackage{amsthm}
\usepackage{amssymb}
\usepackage{mdframed}
\usepackage[ruled,vlined]{algorithm2e}[H]
\usepackage{hyperref}
\theoremstyle{definition}
\newcommand{\comment}[1]{}

\usepackage{mdframed}

\newmdtheoremenv{problem_stmt}{Problem}

\usepackage{cite}
\usepackage{amsmath,amssymb,amsfonts}
\usepackage{algorithmic}
\usepackage{graphicx}
\usepackage{textcomp}
\usepackage{xcolor}
\def\BibTeX{{\rm B\kern-.05em{\sc i\kern-.025em b}\kern-.08em
    T\kern-.1667em\lower.7ex\hbox{E}\kern-.125emX}}
\begin{document}
\bstctlcite{IEEEexample:BSTcontrol}
\title{A Prototype on the Feasibility of Learning Spatial Provenance in XBee and LoRa Networks\\
}

\author{\IEEEauthorblockN{Manish Bansal, \space 
  Pramsu Shrivastava and \space   J. Harshan}
\textit{Indian Institute of Technology Delhi, India}\\
bansal21manish@gmail.com, \space ee1210140@iitd.ac.in, \space jharshan@ee.iitd.ac.in

}
\maketitle

\begin{abstract}
In Vehicle-to-Everything (V2X) networks that involve multi-hop communication, the Road Side Units (RSUs) typically desire to gather the location information of the participating vehicles to provide security and network-diagnostics features. Although Global Positioning System (GPS) based localization is widely used by vehicles for navigation; they may not forward their exact GPS coordinates to the RSUs due to privacy issues. Therefore, to balance the high-localization requirements of RSU and the privacy of the vehicles, we demonstrate a new spatial-provenance framework wherein the vehicles agree to compromise their privacy to a certain extent and share a low-precision variant of its coordinates in agreement with the demands of the RSU. To study the deployment feasibility of the proposed framework in state-of-the-art wireless standards, we propose a testbed of ZigBee and LoRa devices and implement the underlying protocols on their stack using correlated Bloom filters and Rake compression algorithms. Our demonstrations reveal that low-to-moderate precision localization can be achieved in fewer packets, thus making an appealing case for next-generation vehicular networks to include our methods for providing real-time security and network-diagnostics features.  

\end{abstract}


\section{Introduction}


V2X networks, which typically include communication among vehicles and Road Side Units (RSU), are expected to enhance road safety and traffic efficiency as part of smart city initiatives \cite{v2x}. Given that mission-critical data are conveyed through their packets over a wireless medium, V2X networks are also susceptible to cyber-security threats from external adversaries \cite{v2x_attacks}. As a result, next-generation V2X networks should possess the capability to detect such security threats and initiate appropriate mitigation strategies. This demonstration paper focuses on the feasibility of implementing wireless protocols which capture the data-flow logs at the RSUs to detect security threats on V2X networks.\looseness=-1

\subsection{Importance of Spatial-Provenance}

In V2X networks, vehicles may not be able to communicate directly with other vehicles or RSUs either because of transmit-power constraints or signalling blockage-effects. Therefore, vehicles should be connected in such a way that messages from source vehicle are communicated to the RSU with the help of several intermediate vehicles in a multi-hop fashion. In such scenarios, the RSU must be able to remotely learn the state of the network when it receives packets, such as the identity of the packet forwarders, path travelled by the packet \cite{amogh}, \cite{suraj} and \cite{nodeembedding}, the location of the vehicles that forwarded or originated the packet, and other diagnostic parameters. In particular, if the RSU has the knowledge about the vehicles' locations, it can offer location-based security features and other network diagnostic features. \looseness=-1

\subsection{Privacy Issues with Spatial-Provenance}

To assist location-based security features, vehicles may embed their Global Positioning System (GPS) coordinates when forwarding the packet, as GPS is anyway used by vehicles for navigation purposes. However, in a multi-hop communication setup, the participating vehicles may not want to share their exact location since the RSU and the other vehicles can learn their exact location from the packet. Although embedding an encrypted version of their GPS coordinates in the packet is one way to keep their location private from third-party observers, such a process will reveal their exact location to the RSU, and also increase end-to-end delay on the packets.\looseness=-1 

Based on the above discussion, it is clear that while the RSU may want the exact location of the vehicles, the forwarding vehicles may not want to reveal the same. Therefore, to achieve a balance between the RSU's expectations and vehicles' privacy, we present an amicable solution wherein the RSU first divides its coverage area into several fragments of equal size, and then requests the vehicles to embed the identities of their fragments when forwarding the packets. Since the size of each fragment represents the granularity of localization, the vehicles can agree upon a suitable size of the fragments without revealing their exact location. Thus, the choice of the fragment size serves as the underlying parameter for preserving privacy on spatial-provenance.\looseness=-1   

\begin{figure}[ht!]
     \centering
    \includegraphics[trim={0 0 0 0},clip,scale=0.52]{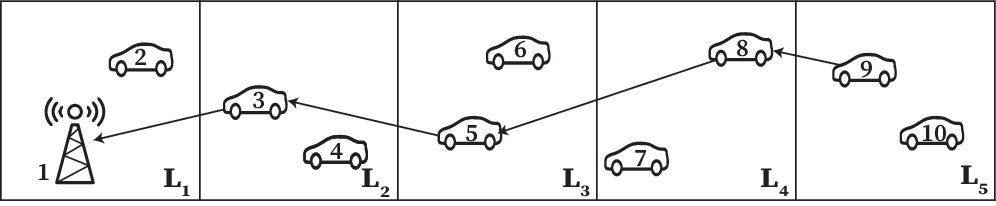}
        \caption{A depiction of area fragmentation wherein the coverage area of an RSU has been divided into 5 segments, and the vehicles are asked to reveal the identity of their segments instead of their exact location.\looseness=-1}
        \label{fig: network Image}
\end{figure}

In the context of vehicular networks, since roads can be linearly modelled, a linear stretch of road in the communication range of RSU can be divided into linear segments of equal length, exemplified in Fig. \ref{fig: network Image}. Subsequently, the RSU can broadcast the segmentation information to vehicles in the form of a dictionary. Finally, with the help of the dictionary received from RSU, the vehicles can learn the identity their segment using their own GPS coordinates, and then embed the same information in their packets. In the next section, we discuss implementation aspects of the above framework and also highlight the need for demonstrating the same on state-of-the-art wireless protocols.\looseness=-1

\subsection{Impact of this Testbed}

In order to implement the proposed framework, we use space-efficient Bloom filters as the underlying data structures to carry the information on spatial-provenance. As a consequence, each vehicle will embed its segment identity on a dedicated set of shared bits in the packet before forwarding it to the next vehicle, en-route to the RSU. This process facilitates the RSU to recover the segment identity of each vehicle through a simple constant complexity verification-task. While the idea of a Bloom-filter based solution for the proposed framework is appealing, it is not clear whether it is feasible to implement such strategies on a state-of-the-art payload constrained wireless protocols such as ZigBee and LoRa. Specifically, some of the underlying questions regarding the feasibility of implementation are \looseness=-1
\begin{enumerate}
\item What is the percentage of the payload space of ZigBee and LoRa occupied for spatial-provenance as a function of the number of vehicles, number of segments, and the reliability of provenance recovery?  
\item What is the end-to-end delay overhead in implementing the proposed spatial-provenance framework on ZigBee and LoRa networks? 
\end{enumerate}
To answer the above questions, building a wireless prototype with spatial-provenance is important as it will enable us to measure several metrics related to space- and time-complexity, and also study the granularity with which the RSU can learn spatial-provenance of the vehicles when using ZigBee or LoRa. Furthermore, this prototype will also help us in recommending the proposed spatial-provenance framework for implementation to standardization bodies such as IEEE, ZigBee consortium and LoRa alliance.\looseness=-1 

\begin{figure}[!h]
    \centering
    \includegraphics[width = 0.49\textwidth]{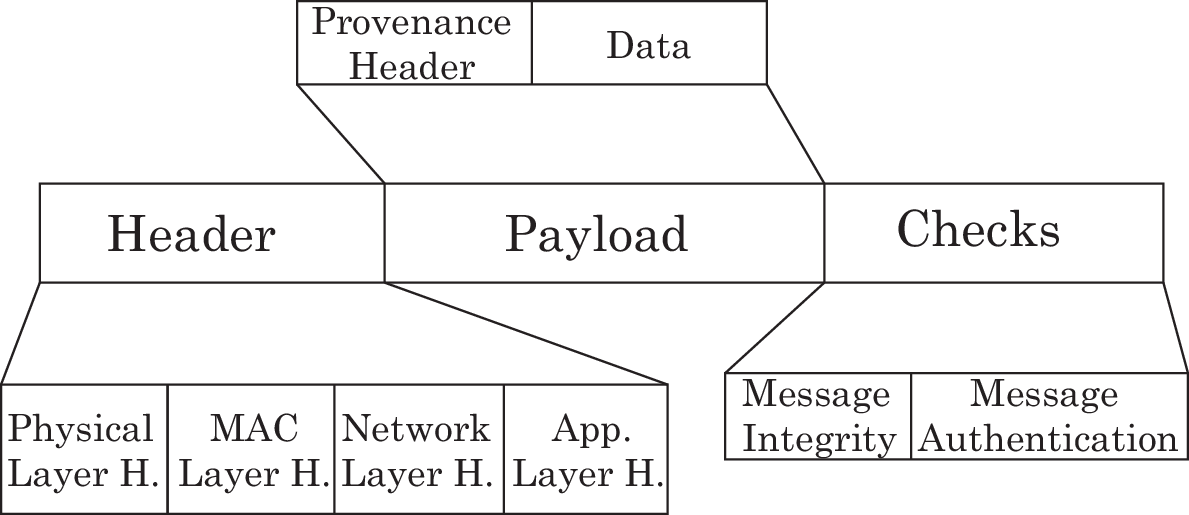}
    \setlength{\belowcaptionskip}{-10pt} 
    \caption{Depiction of general structure of a network packet, where H. implies Header. In our testbed, part of the payload will be used to convey the spatial-provenance information. \looseness=-1}
    \label{fig:Payload}
\end{figure}

In the next section, we present the implementation aspects of our wireless prototype. Since there is no practical implementation of spatial-provenance framework for existing wireless networks, no dedicated bits are allocated for the same in their packet structure. Therefore, in our prototype, we use a portion of the payload to carry the Bloom filter for spatial-provenance, as shown in Fig. \ref{fig:Payload}. \looseness=-1




\section{Testbed Setup for Spatial-provenance }

\begin{figure}[!h]
    \centering
    \includegraphics[width = 0.49\textwidth]{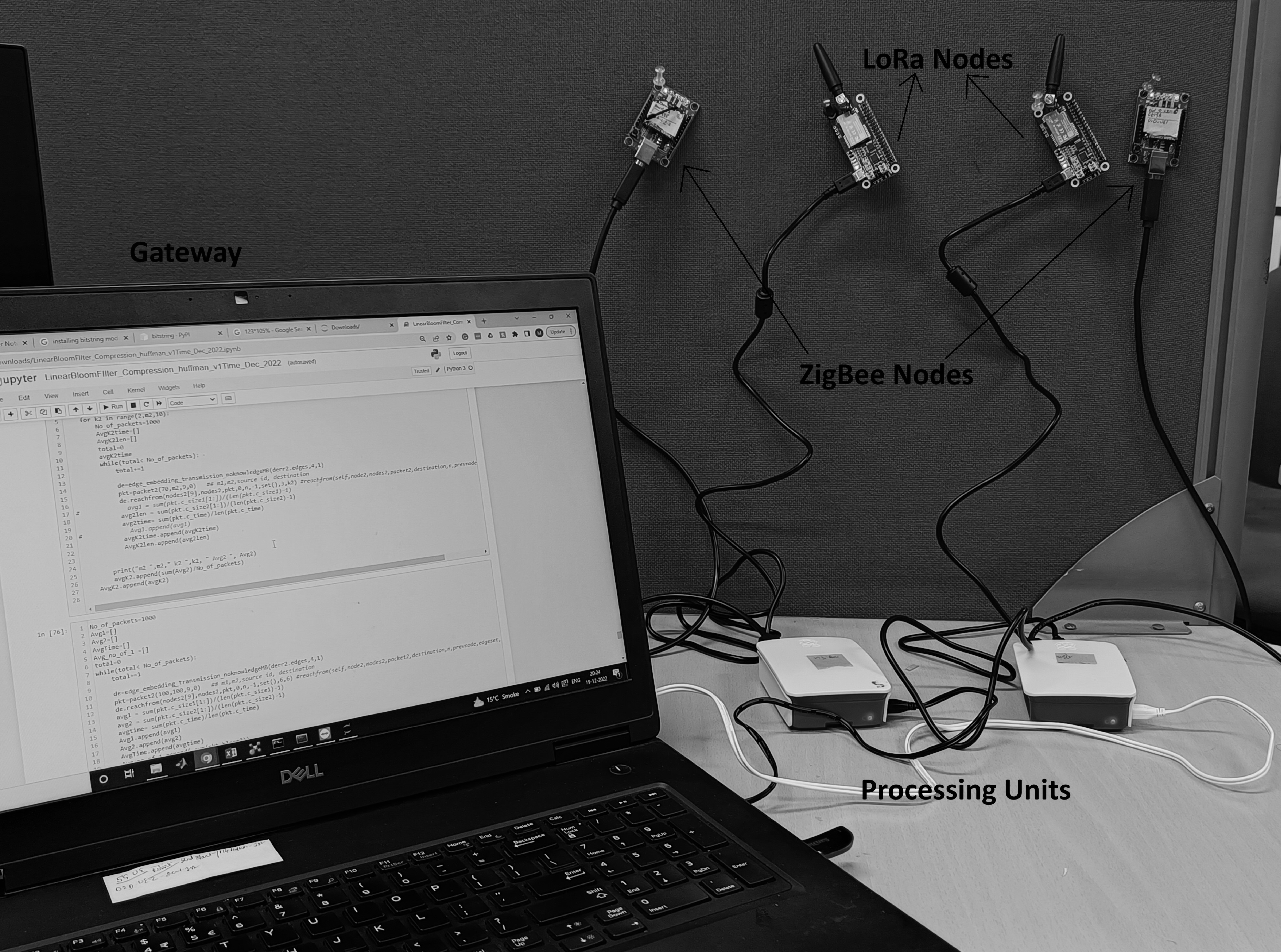}
    \setlength{\belowcaptionskip}{-10pt} 
    \caption{Testbed comprising XBee and LoRa devices to demonstrate a static vehicular network.\looseness=-1}
    \label{fig:XBee}
\end{figure}

As shown in Fig. \ref{fig:XBee}, our testbed setup consists of XBee S2C devices, which work on ZigBee protocol in the ISM band of 2.4 GHz with 16 channels and a bandwidth of 5 MHz. For long-range communication, we use the LoRa modules manufactured by Semtech, which works at 868 MHz in India. For computing purposes, we use Raspberry Pis as well as laptops. We model a vehicle using a combination of Raspberry Pi and XBee S2C devices to demonstrate a static vehicular network with short-range communication. Whereas, for long-range communication, we model a vehicle using LoRa and Raspberry Pi. In both these cases, a high-performance computing device plays the role of RSUs and either an XBee device or LoRa.\looseness=-1  


\subsection{Hardware Setup}
To prepare the setup, we distribute the vehicular nodes across a geographical area such that any node can communicate with the RSU in a multi-hop manner. Given that vehicles are static in our setup, we hardcode their location information on them so that they can identify their segment identity upon receiving the dictionary from the RSU. For the routing protocol AODV is used for multi-hop communication from every vehicle to RSU. For downlink communication, RSU communicates directly with every vehicle in single hop. \looseness=-1
\subsection{Message Flow for Spatial-provenance}
Once the vehicles are registered with the RSU and have completed the neighbour discovery protocol, RSU initiates the broadcast phase, wherein, the dictionary comprising the number of segments and their identities is communicated to all vehicles. We highlight that the chosen number of segments is apriori decided in mutual agreement with the vehicles to preserve their privacy.\looseness =-1

\begin{figure}[ht!]
     \centering
    \includegraphics[trim={0 0 0 0},clip,scale=0.45]{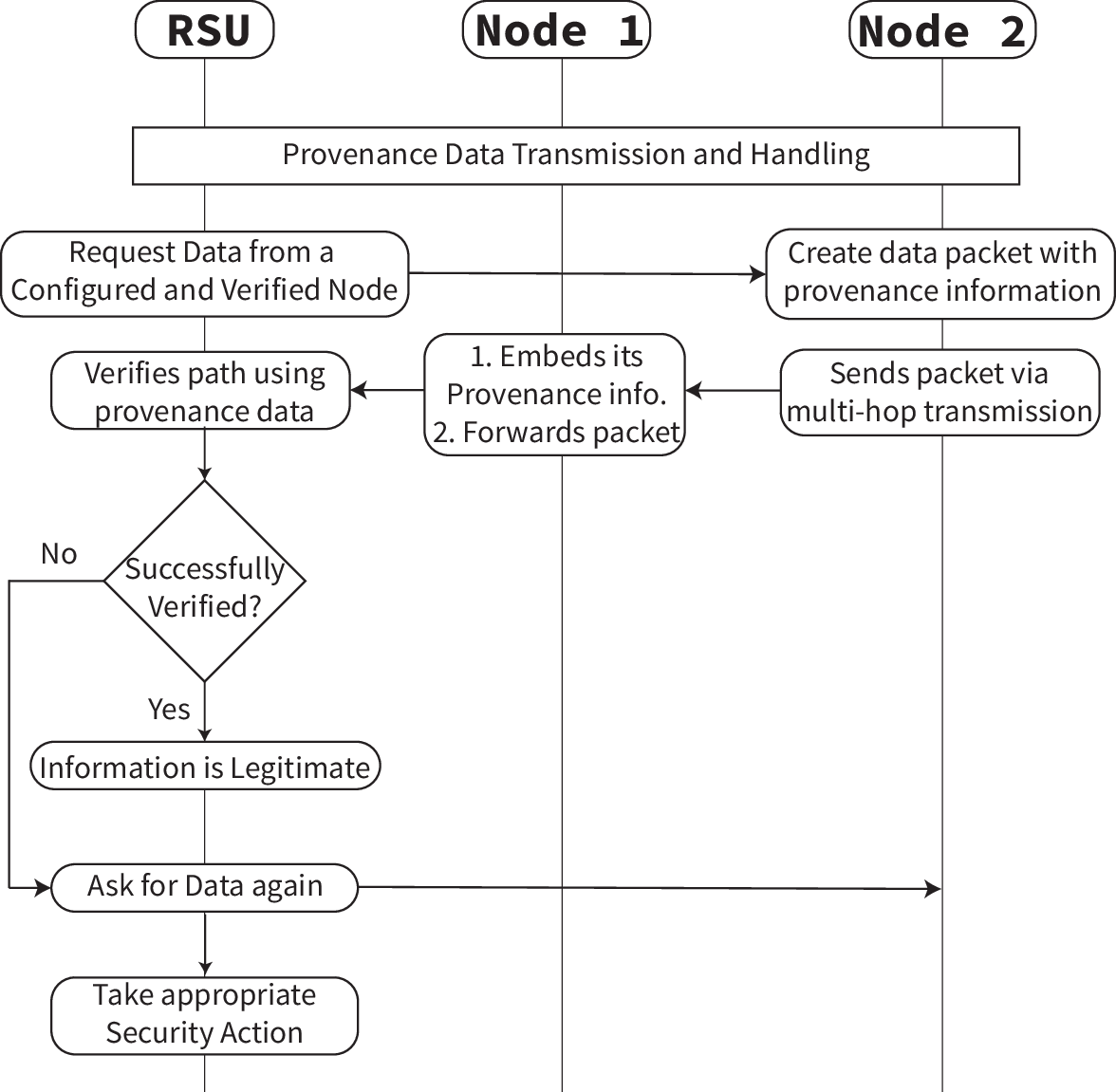}
        \setlength{\belowcaptionskip}{-10pt}   
        \caption{Flow of messages between RSU and nodes for conveying spatial-provenance in a two-hop network.\looseness=-1}
        \label{fig: steps_diagram_2}
\end{figure}

As exemplified in Fig. \ref{fig: steps_diagram_2}, a source vehicle embeds its spatial-provenance into the Bloom filter of the packet and forwards it to the next vehicle in the path. Subsequently, the next vehicle repeats the process of embedding its spatial-provenance until the packet reaches the RSU. To execute these steps, we choose the underlying parameters such as number of hash function for Bloom filter and the Bloom filter size, based on an offline optimization process. Furthermore, given that Bloom filters are inherently sparse, i.e., having fewer number of ones than zeros in their data structure, we ask each vehicle to compress its provenance using RAKE compression \cite{rake}. Consequently, every vehicle that intends to embed its spatial-provenance implements a RAKE decompression algorithm on the reception of a packet. Finally, once the packet is received at the RSU, it verifies the location of the vehicles using their identities and the dictionary. This way, every vehicle is localised at the RSU, respecting the privacy constraints of the vehicles. \looseness=-1

\section{Takeaways from our Testbed}

We conducted several experiments on our testbed and recorded the space- and time-complexity metrics of our framework for a 5-hop network of 10 vehicles spread across 5 location segments with each segment having 2 vehicles. Bloom filter size of 100 bits are used by the vehicles along with 8 hash functions to embed the identities of their segments. Since Bloom filters are sparse, we measure its average sparsity, which is defined as the ratio of the sum of the number of bits lit in a Bloom filter across all hops by the product of the number of hops and the Bloom filter size. After implementing the RAKE compression algorithm at each hop, we also measure the average provenance size, which is calculated as the ratio of the sum of the size of compressed provenance at each node to the number of hops. With ten thousand iterations of experiments, the above metrics are listed in Table \ref{tab:compression_table} for Bloom filter sizes of 100, 125 and 150. For instance, with 100 bits for Bloom filter, the average sparsity will be $21\%$, which can be compressed to $~77$ bits. Therefore, when using XBee, which has a fixed payload of 255 bytes, our spatial-provenance needs only $3.9\%$ of the payload. Similarly, when using LoRa (data rate 7), which has a fixed payload of 222 bytes, our spatial-provenance occupies $4.5\%$ of the payload. In contrast, without compression, our spatial-provenance would occupy roughly $\sim6\%$ of the payload in both XBee and LoRa. While RAKE compression reduces the space, we observe that it increases the end-to-end packet delay compared to the scheme without any compression algorithm.

As the LoRa network is used for long-range communication, the coverage area is much larger, whereas the XBee coverage area is smaller due to short-range communication. Therefore, for the same number of segments, absolute privacy is more in LoRa and less in XBee; however, the normalized privacy is the same in both technologies. Overall, our testbed proves that spatial-provenance algorithms can be accommodated on ZigBee and LoRa as long as the RSU needs to learn low-to-moderate resolution of localization. However, a higher resolution of localization is possible only with a larger percentage of payload space.\looseness=-1 
\begin{table}[!h]
    \centering
    \caption{Compressed provenance size}
    \begin{tabular}{|c|c|c|}
    \hline
         Provenance size  & Sparsity (Avg.)   & Avg. Provenance size  \\(bits) & $\%$ & after compression (bits)  \\
    \hline
         100 & 20.92 & 76.92 \\
         125 & 17.17 & 86.34\\
         150 & 14.58 & 94.03 \\
    \hline
    \end{tabular}
    \label{tab:compression_table}
\end{table}
\section*{Acknowledgement}
This work was supported by the CARS project titled ``Development of Network Provenance Techniques for Monitoring Wireless Networks" from the DRDO, India.\looseness=-1

\looseness=-1

\end{document}